\documentstyle[12pt]{article}
\begin{document}
\input{amssym.tex}

\title{Polarized Dirac fermions in de Sitter spacetime}

\author{Ion I. Cot\u aescu\\ {\it  West University of Timi\c soara,}\\{\it V. 
Parvan Ave. 4, RO-1900 Timi\c soara}}

\maketitle

\begin{abstract}
The tetrad gauge invariant theory of the free Dirac field in two special 
moving  charts of the de Sitter spacetime is investigated pointing out 
the operators that commute with the Dirac one. These are the generators of 
the symmetry transformations corresponding to isometries that give rise to 
conserved quantities according to the Noether theorem. With their help the 
plane wave spinor solutions of the Dirac equation with given momentum and 
helicity are derived and the final form of the quantum Dirac field is 
established. It is shown that the canonical quantization  leads to a correct 
physical interpretation of the massive or massless fermion quantum fields.  

Pacs: 04.62.+v
\end{abstract}
\   

\newpage
\section{Introduction}

The recent astrophysical investigations showing that the expansion of the
universe is accelerating  \cite{TR} may increase the interest for the de 
Sitter spacetime which could represent the far future limit of the actual 
universe. On the other hand, the Dirac fermions (leptons and quarks) are the 
principal components of the matter since their gauge symmetries determine the 
main features of the physical picture. For these reasons, we believe that the 
study of the tetrad gauge invariant theory of the free Dirac field in de Sitter 
background may be important for understanding the influence of the external 
gravitational field minimally coupled with the fermion fields.     

The Dirac equation in de Sitter spacetime (of radius $R=1/\omega=
\sqrt{3/\Lambda_c}$, produced by the cosmological constant $\Lambda _c$) was 
studied in moving or static local charts (i.e., natural frames) suitable for 
separation of variables that should lead to significant analytical solutions 
\cite{SHI,C1}. In \cite{SHI} spherical wave solutions were derived in the 
local chart $\{t,r,\theta,\phi\}$ commonly associated to the Cartesian one 
$\{t, \vec{x}\}$ having the line element 
\begin{equation}\label{mssu}
ds^{2}=dt^2 - e^{2\omega t} d\vec{x}^2\,, 
\end{equation}
in which the form of the plane wave solutions in a Cartesian tetrad gauge was 
outlined. Since in these moving charts the operator $i\partial_{t}$ is 
no longer a Killing vector field, the quantum modes corresponding to these 
solutions have no well-determined energies. Obviously, this is not an 
impediment but, in addition, there are some integration constants the physical 
meaning of which remains obscure. An alternative might be the particular 
solutions obtained through the separation of variables in static charts since 
these are energy eigenspinors. Recently, we found such solutions in a static 
central chart with the help of another version of Cartesian tetrad gauge 
\cite{C1}. These have integration constants with precise physical significance 
but, unfortunately, their form is too complicated to be normalized in the 
energy scale. Therefore, actually we do not have yet a complete set of 
particular solutions that may be used for writing the general form of the 
quantum field \cite{SW1}.
     
Looking for a such set of normalized particular solutions, we 
continue here the investigations of the free Dirac field in the chart with the 
line element (\ref{mssu}). Our aim is to write down the plane wave solutions 
suggested in \cite{SHI}, defined as common eigenspinors of a complete set of 
commuting observables whose eigenvalues should determine the constants arising 
from the separation of variables. The main purpose of the present article is 
to show that these solutions are suitable for expressing the canonically 
quantized Dirac field in terms of creation and annihilation operators of 
fermions with good physical properties.     

To this end we exploit the results of our theory of external symmetry \cite{C2} 
which explains the relations among the geometric symmetries and the operators 
commuting with the Dirac one written with the help of the Killing vectors 
long time ago \cite{CML}. In fact these operators are nothing other than the 
generators of the spinor representation of the universal covering group of the 
isometry group \cite{C2} and, therefore, they represent the main physical 
observables among which we can choose different sets of commuting operators 
defining quantum modes. This method is efficient especially in the case of the 
de Sitter spacetime where the high symmetry given by the $SO(4,1)$ isometry 
group \cite{SW,WALD} offers one the opportunity of a very rich algebra of 
operators able to get physical meaning. 

Our basic idea is that the significance of these observables is independent on 
the choice of the local chart and tetrad gauge, even though their form is 
strongly dependent on both these elements. For this reason we interpret the 
generators of the subgroup $E(3)\subset SO(4,1)$ as the three-dimensional 
momentum and angular momentum operators and take over the Hamiltonian operator 
found already in static frames \cite{C1,C2}. From this algebra we select the 
momentum operators and, in addition, we construct an one-component 
Pauli-Lubanski (or helicity) operator \cite{WKT} obtaining thus the set of 
commuting observables that defines quantum modes with given momentum and 
helicity. We show that the common eigenspinors of these operators are the 
desired plane wave solutions of the Dirac equation which can be easily 
normalized in the momentum scale. Moreover, we demonstrate that the set of 
these solutions is complete (in generalized sense).

This set is used for expanding the free Dirac field in terms of creation and 
annihilation operators of fermions characterized by momentum and helicity, 
pointing out that the canonical quantization requires to adopt the standard 
anticommutation rules in momentum representation. In this way the conserved 
quantities predicted by the Noether theorem become the one-particle operators 
of the quantum field theory among them the diagonal ones are the momentum, 
helicity and charge operators. The Hamiltonian operator is not diagonal in this 
basis since it does not commute with the momentum components. The conclusion is 
that in our approach the second quantization can be done in canonical manner 
obtaining new results specific to the de Sitter geometry.  However, the free 
fermions on this background have some properties similar to those found in 
Minkowski spacetime as, for example, the well-known law of neutrino 
polarization \cite{SW1}. 

We start in the second section with a brief review of the main results of our 
theory of external symmetry presenting the form of the symmetry generators and 
the conservation laws due to the Noether theorem. The moving chart we use here 
are introduced in section 3 where, in addition, we define a suitable basis 
of the $SO(4,1)$ generators helping us to identify  the momentum, angular 
momentum and Hamiltonian operators. The plane wave solutions of the Dirac 
equation with fixed momentum and helicity are written in the next section 
while the section 5 is devoted to the canonical quantization of the Dirac 
field.  After we present our concluding remarks, in an appendix we give 
the form of the isometries in the charts and parametrization used here and 
in another we give few useful formulas.

We use natural units with $\hbar=c=1$.

\section{Gauge and external symmetry}

In curved spacetimes, $M$, the choice of the local charts, 
$\{x\}$, of coordinates $x^{\mu}$ ($\mu, \nu,...= 0,1,2,3$) is important from 
the observer point of view. Moreover, the tetrad gauge covariant theory of the 
fields with spin requires to explicitly use the tetrad fields, $e_{\hat\mu}(x)$ 
and $\hat e^{\hat\mu}(x)$, giving the local frames and the corresponding 
coframes. These are labeled by the local indices, 
$\hat\mu, \hat\nu,...=0,1,2,3$, and have the orthonormalization properties   
$e_{\hat\mu}\cdot e_{\hat\nu}=\eta_{\hat\mu \hat\nu}$, 
$\hat e^{\hat\mu}\cdot \hat e^{\hat\nu}=\eta^{\hat\mu \hat\nu}$, and 
$\hat e^{\hat\mu}\cdot e_{\hat\nu}=\delta^{\hat\mu}_{\hat\nu}$, 
with respect to the Minkowski metric $\eta=$diag$(1,-1,-1,-1)$. The 1-forms,  
$d\hat x^{\hat\mu}=\hat e_{\nu}^{\hat\mu}dx^{\nu}$, allow one to write 
the line element  
\begin{equation}\label{(met)}
ds^{2}=\eta_{\hat\mu \hat\nu}d\hat x^{\hat\mu}d\hat x^{\hat\nu}=
g_{\mu \nu}(x)dx^{\mu}dx^{\nu}
\end{equation}   
defining the metric tensor $g_{\mu \nu}$ which raises or lowers the Greek 
indices 
while for the hated Greek ones we have to use the Minkowski metric. The 
derivatives in local frames are the vector 
fields  $\hat\partial_{\hat\nu}=e^{\mu}_{\hat\nu}\partial_{\mu}$ which satisfy 
the commutation rules  $[\hat\partial_{\hat\mu},\hat\partial_{\hat\nu}]
=C_{\hat\mu \hat\nu 
\cdot}^{~\cdot \cdot \hat\sigma}\hat\partial_{\hat\sigma}$
giving the Cartan coefficients that help us to write the conecttion 
components in local frames.

Let $\psi$ be a Dirac free field of  mass $m$, defined on the  space domain 
$D$, and $\overline{\psi}=\psi^{+}\gamma^{0}$ its Dirac adjoint.  The 
tetrad gauge invariant action of the Dirac field minimally coupled with the 
gravitational field is      
\begin{equation}\label{action}
{\cal S}[e,\psi]=\int\, d^{4}x\sqrt{g}\left\{
\frac{i}{2}[\overline{\psi}\gamma^{\hat\alpha}D_{\hat\alpha}\psi-
(\overline{D_{\hat\alpha}\psi})\gamma^{\hat\alpha}\psi] - 
m\overline{\psi}\psi\right\}
\end{equation}
where  $g=|\det(g_{\mu\nu})|$ while $\gamma^{\hat\alpha}$ are the Dirac matrices 
which accomplish
$\{ \gamma^{\hat\alpha},\, \gamma^{\hat\beta} \}=
2\eta^{\hat\alpha \hat\beta}$. 
The covariant derivatives in local frames,
$D_{\hat\alpha}=\hat\partial_{\hat\alpha}+\hat\Gamma_{\hat\alpha}$,
are expressed in terms of the spin connections
\begin{equation}
\hat\Gamma_{\hat\mu}=
\hat\Gamma_{\hat\mu\hat\nu\hat\lambda}S^{\hat\nu \hat\lambda}=
\frac{i}{4}(C_{\hat\mu \hat\nu \hat\lambda}-
C_{\hat\mu \hat\lambda \hat\nu}-C_{\hat\nu \hat\lambda \hat\mu})
S^{\hat\nu\hat\lambda}
\end{equation}
given by the basis-generators in covariant parametrization, 
$S^{\hat\alpha \hat\beta}=i [\gamma^{\hat\alpha}, \gamma^{\hat\beta} ]/4$, 
of the reducible spinor representation $ \rho\sim (1/2,0)\oplus (0,1/2)$ of the 
$SL(2,\Bbb C)$ group \cite{WKT,TH} (i.e., the universal covering group of the 
Lorentz group, $L^{\uparrow}_{+}$, that is the gauge group of the metric $\eta$).   
The Dirac operator of the equation $E_{D}\psi=0$ derived from (\ref{action}) 
reads $E_{D}=i\gamma^{\hat\alpha}D_{\hat\alpha} - m$.
In other respects, from the conservation of the electric charge one  
deduces that when $e^{0}_{i}=0$  ($i,\,j,...=1,2,3$) then the time-independent 
relativistic scalar product of two spinors \cite{BD},
\begin{equation}\label{sp}
\left< \psi,\psi^{\prime}\right>=\int_{D}d^{3}x\,\mu(x)
\overline{\psi}(x)\gamma^{0}\psi^{\prime}(x)  \,, 
\end{equation}
has the weight function  $\mu=\sqrt{g}\,e_{0}^{0}$.

The action (\ref{action}) is gauge invariant in the sense that 
it remains unchanged when one performs a gauge transformation 
\begin{eqnarray}
\psi(x) &\to& \psi^{\prime}(x)=\rho[A(x)]\psi(x) \\ 
e_{\hat\alpha}(x)&\to& e^{\prime}_{\hat\alpha}(x)=\Lambda_{\hat\alpha\,\cdot}
^{\cdot\,\hat\beta}[A(x)]e_{\hat\beta}(x)
\end{eqnarray}
produced by $A(x)\in SL(2,\Bbb C)$  and $\Lambda[A(x)] \in L^{\uparrow}_{+}$.
Based on this symmetry, we have defined the group of external symmetry, 
$S(M)$, corresponding to the isometry group $I(M)$. 
The transformations of $S(M)$ are isometries of $I(M)$, 
$x\to x^{\prime}=\phi_{\xi}(x)$ (depending on the  parameters $\xi^a$, $a=1,2,...$), 
combined with appropriate gauge transformations in such a manner 
to preserve the tetrad gauge. 
In a fixed gauge, one associates to each isometry $\phi_{\xi}$ the section 
$A_{\xi}(x)\in SL(2,\Bbb C)$ defined by  
\begin{equation}\label{Axx}
\Lambda^{\hat\alpha\,\cdot}_{\cdot\,\hat\beta}[A_{\xi}(x)]=
\hat e_{\mu}^{\hat\alpha}[\phi_{\xi}(x)]\frac{\partial \phi^{\mu}_{\xi}(x)}
{\partial x^{\nu}}\,e^{\nu}_{\hat\beta}(x) 
\end{equation}    
with the supplementary condition $A_{\xi=0}(x)=1\in SL(2, C)$. 
Then the transformations of $S(M)$  are
\begin{equation}\label{es}
(A_{\xi},\phi_{\xi}):\qquad
\begin{array}{rlrcl}
x&\to&x^{\prime}&=&\phi_{\xi}(x)\\
e(x)&\to&e^{\prime}(x^{\prime})&=&e[\phi_{\xi}(x)]\\
\hat e(x)&\to&\hat e^{\prime}(x^{\prime})&=&\hat e[\phi_{\xi}(x)]\\
\psi(x)&\to&\psi^{\prime}(x^{\prime})&=&\rho[A_{\xi}(x)]\psi(x)\,.
\label{rpsi}
\end{array}
\qquad
\end{equation}
In \cite{C2} we presented arguments that $S(M)$ is the universal covering group 
of $I(M)$ but the representation defined by the last of equations (\ref{rpsi}) 
is not an usual linear representation of $S(M)$ since it is {\em induced} by 
the representation $\rho$ of the group $SL(2,\Bbb C)$ which, in general, 
differs from $S(M)$. For this reason we  say that $\psi$ transforms according 
to the spinor representation of $S(M)$ induced by $\rho$.

The transformations (\ref{rpsi}) leave invariant the form of the  operator 
$E_{D}$ in local frames.  Consequently, each Killing vector, 
$k_{a}=(\partial_{\xi^{a}}\phi_{\xi})_{|\xi=0}$, defines a  basis-generator  
of the spinor representation \cite{C2},
\begin{equation}\label{Xcov}
X_{a}=-ik^{\mu}_{a}D_{\mu}+\frac{1}{2}\,
k_{a\, \mu;\nu}\,e^{\mu}_{\hat\alpha}\,e^{\nu}_{\hat\beta}\,
S^{\hat\alpha\hat\beta}\,, 
\end{equation}
which {\em commutes} with $E_{D}$. (Here the notation $~_{;\nu}$ stands for 
the usual covariant derivatives.)   We must specify that this important 
result was obtained for the Dirac field in \cite{CML} without taking 
into account symmetry transformations. In \cite{C2} we have shown
that the generators (\ref{Xcov}) satisfy the commutation relations 
\begin{equation}
[X_a , X_b ]=ic_{abc}X_c\,, \quad a,b,c=1,2,...n
\end{equation}
given by the structure constants of $I(M)$. On the other hand,  
each generator can be split in orbital and  spin parts as 
$X_{a} = L_{a}+S_{a}$  where the orbital terms, 
\begin{equation}\label{La}
L_a=-ik_{a}^{\mu}(x)\partial_{\mu}\,, 
\end{equation}
are the basis-generators of the natural representation of $I(M)$ carried by 
the space of scalar functions over $M$. The spin terms
\begin{equation}\label{Sx}
S_{a}(x)=
\frac{1}{2}\,\Omega^{\hat\alpha\hat\beta}_{a}(x)S_{\hat\alpha\hat\beta} 
\end{equation}
are defined with the help of the functions 
\begin{equation}\label{Om}     
\Omega^{\hat\alpha\hat\beta}_{a}
=\left( \hat e^{\hat\alpha}_{\mu}\,\partial_{\nu} k_{a}^{\mu}
+k_{a}^{\mu}\partial_{\mu}\hat e^{\hat\alpha}_{\nu}\right)
e^{\nu}_{\hat\lambda}\eta^{\hat\lambda\hat\beta}
\end{equation}
that are antisymmetric if and only if $k_{a}$ is a Killing vector. Thus we 
see that the spin terms of the generators $X_{a}$ generally depend on $x$ and, 
therefore, they do not commute with the orbital terms. However, when $L_{a}$ 
and $S_{a}$ commute between themselves we say that the Dirac field transforms 
manifestly covariant under the transformations of the subgroup parametrized by  
$\xi^a$.

Our theory of external symmetry offers us the framework we need to calculate 
the conserved quantities predicted by the Noether theorem. Starting with the 
infinitesimal transformations of the one-parameter subgroup of $S(M)$ 
generated by $X_a$, we find that there exists the current 
$\Theta^{\mu}[X_a]$ which satisfies $\Theta^{\mu}[X_a]_{;\mu}=0$. For the 
action (\ref{action}) this is  
\begin{equation}
\Theta^{\mu}[X_a]=-\tilde T^{\mu\,\cdot}_{\cdot\,\nu}k_a^{\nu}+
\frac{1}{4}  
\,\overline{\psi}\{\gamma^{\hat\alpha}, S^{\hat\beta \hat\gamma} \}\psi\,
e^{\mu}_{\hat\alpha} \,\Omega_{a\,\hat\beta \hat\gamma}
\end{equation}
where
\begin{equation}
\tilde T^{\mu\,\cdot}_{\cdot\,\nu}=
\frac{i}{2}\left[\overline{\psi}\gamma^{\hat\alpha}e^{\mu}_{\hat\alpha}
\partial_{\nu}\psi-
(\overline{\partial_{\nu}\psi})\gamma^{\hat\alpha}e^{\mu}_{\hat\alpha}\psi
\right]
\end{equation}
is a notation for a part of the stress-energy tensor of the Dirac field 
\cite{SW,BD}. Finally, it is clear that the corresponding conserved quantity 
is the real number
\begin{equation}\label{cq}
\int_{D} d^3 x \sqrt{g}\,\Theta^{0}[X_a]= 
\frac{1}{2}\left[\left<\psi, X_a\psi\right>+\left<X_a\psi, \psi\right>\right]\,.
\end{equation}
We note that it is premature to interpret this formula  as an expectation 
value or to speak about Hermitian conjugation of the operators $X_a$ with 
respect to (\ref{sp}) before to specify the boundary conditions on $D$. 
However, what is important here is that (\ref{cq}) is useful in quantization 
giving directly the one-particle operators of the quantum field theory. 

\section{Observables in de Sitter spacetime}

Let us consider now  $M$ be the de Sitter spacetime defined as the 
hyperboloid of radius $R$ in the five-dimensional flat spacetime $M^5$ of 
coordinates $Z^A$ $(A,\,B,...= 0,1,2,3,5)$ and metric 
$\eta^5={\rm diag}(1,-1,-1,-1.-1)$ \cite{SW}. The hyperboloid equation,  
\begin{equation}\label{hip}
\eta^5_{AB}Z^A Z^B=-R^2\,,  
\end{equation}
defines $M$ as an homogeneous space of the pseudo-orthogonal group $SO(4,1)$  
which is in the same time the  gauge group of the metric $\eta^{5}$ and the 
isometry group, $I(M)$, of the de Sitter spacetime. For this reason it is 
convenient to use the covariant  real parameters 
$\xi^{AB}=-\xi^{BA}$ since then the orbital basis-generators of the natural 
representation of $SO(4,1)$, carried by the space of scalar functions over 
$M^{5}$, have the standard form
\begin{equation}\label{LAB5}
 L_{AB}^{5}=i\left[\eta_{AC}^5 Z^{C}\partial_{B}-
 \eta_{BC}^5 Z^{C}
\partial_{A}\right].
\end{equation} 
They  will give us directly the orbital basis-generators, $L_{(AB)}$, of the 
scalar representations of  $I(M)$. Indeed, if one knows the 
functions $Z^{A}(x)$  solving (\ref{hip}) in the chart $\{x\}$ then one can 
write (\ref{LAB5}) in the form  (\ref{La}) finding the generators $L_{(AB)}$ 
and implicitly the components $k^{\mu}_{(AB)}(x)$ of the 
Killing vectors associated to the parameters $\xi^{AB}$ \cite{C2}. Furthermore, 
one has to calculate the spin parts $S_{(AB)}$, according to (\ref{Sx}) and 
(\ref{Om}), obtaining  the final form of the basis-generators 
$X_{(AB)}=L_{(AB)}+S_{(AB)}$ of the spinor representation of $S(M)$ induced by 
$\rho$.     
   
In the de Sitter spacetime there are many static or moving charts of physical 
interest. Among the moving ones a special role  plays the chart 
$\{t_c, \vec{x}\}$ with the conformal time $t_c$ and Cartesian spaces 
coordinates $x^i$ defined by   
\begin{eqnarray}
Z^0&=&-\frac{1}{2\omega^2 t_c}\left[1-\omega^2({t_c}^2 - r^2)\right] 
\nonumber\\ 
Z^5&=&-\frac{1}{2\omega^2 t_c}\left[1+\omega^2({t_c}^2 - r^2)\right] 
\label{Zx}\\ 
Z^i&=&-\frac{1}{\omega t_c}x^i \nonumber
\end{eqnarray}
with $r=|\vec{x}|$. Even if this chart cover only a half of the manifold $M$, 
for $t_c\in (-\infty,\,0)$  and $\vec{x}\in D\equiv {\Bbb R}^3$, it has the 
advantage of a simple conformal flat line element \cite{BD},  
\begin{equation}\label{mconf}
ds^{2}=\frac{1}{\omega^2 {t_c}^2}\left({dt_c}^{2}-d\vec{x}^2\right)\,.
\end{equation}
Moreover, the conformal time $t_c$ is related through
\begin{equation}
\omega t_c =-e^{-\omega t}
\end{equation} 
to the proper time $t \in (-\infty, \infty)$, of an observer at the point 
$\vec{x}=0$ of the chart $\{t, \vec{x}\}$ with the line element (\ref{mssu}).   
In what follows we study the Dirac field in the chart $\{t,\vec{x}\}$  
using the conformal time as a helpful auxiliary  ingredient. The form of the 
line element (\ref{mconf}) allows one to choose the simple Cartesian gauge 
with  non-vanishing tetrad components \cite{SHI} 
\begin{equation}\label{tt}
e^{0}_{0}=-\omega t_{c}\,, \quad e^{i}_{j}=-\delta^{i}_{j}\,\omega t_c 
\,,\quad
\hat e^{0}_{0}=-\frac{1}{\omega t_{c}}\,, \quad \hat e^{i}_{j}=-\delta^{i}_{j}\,
\frac{1}{\omega t_c}
\end{equation}
in which the Dirac operator reads
\begin{eqnarray}\label{ED1}
E_D&=&-i\omega t_c\left(\gamma^0\partial_{t_{c}}+\gamma^i\partial_i\right)
+\frac{3i\omega}{2}\gamma^{0}-m \nonumber\\ 
&=&i\gamma^0\partial_{t}+ie^{-\omega t}\gamma^i\partial_i
+\frac{3i\omega}{2}\gamma^{0}-m  
\end{eqnarray}
and the weight function of the scalar product (\ref{sp}) is
\begin{equation}\label{mu}
\mu=(-\omega t_{c})^{-3}=e^{3\omega t}\,.
\end{equation}

The next step is to calculate the basis-generators $X_{(AB)}$ of the spinor 
representation of $S(M)$  in this gauge since, as mentioned, these commute with 
$E_{D}$. We observe that the group $SO(4,1)$ includes the subgroup 
$E(3)=T(3)\circledS SO(3)$ of the isometries of the 3-dimensional Eucidean 
space (formed by $\Bbb R^3$ translations and proper  rotations \cite{WKT}) and, 
consequently, its universal covering group, 
$\tilde E(3)=T(3)\circledS SU(2)$, is a subgroup of $S(M)$. The advantage of 
$\tilde E(3)$  is that its generators have the usual physical interpretation 
of  momentum and total angular momentum operators. The problem of the 
Hamiltonian operator seems to be more complicated but we know that in the 
static central charts with the static time $t_{s}$ this is  
$H=\omega X_{(05)}=i\partial_{t_{s}}$ \cite{C2}. Thus the Hamiltonian 
operator and the components of momentum, $\vec{P}$, and total angular 
momentum, $\vec{J}$ ($J^i=\varepsilon_{ijk}J_{jk}/2$), operators can be 
defined as 
\begin{eqnarray}
H&\equiv&\omega X_{(05)}=-i\omega(t_{c}\partial_{t_{c}}+x^{i}\partial_i)
\label{Gi}\\
P^{i}&\equiv&\omega\left(X_{(5i)}-X_{(0i)}\right)=-i\partial_{i}\\
J_{ij}&\equiv&X_{(ij)}=-i(x^i\partial_j-x^j\partial_i)+S_{ij} 
\end{eqnarray}
remaining with the generators 
\begin{equation}\label{Gf}
N^i\equiv X_{(5i)}+X_{(0i)}=\omega ({t_{c}}^2-
r^2)P^i + 2 x^i H+2\omega( S_{i0}t_{c}+ S_{ij}x^j)
\end{equation}
which do not have an obvious physical significance. The $SO(4,1)$ 
transformations corresponding to these generators and the associated 
isometries of the chart $\{t_{c}, \vec{x}\}$ are briefly presented 
in the Appendix A.

In the other local chart, $\{t,\vec{x}\}$, the Hamiltonian operator reads 
\begin{equation}\label{Ham}
H=i\partial_{t}+\omega\, \vec{x}\cdot\vec{P}
\end{equation}       
while the operators $\vec{P}$ and $\vec{J}=\vec{L}+\vec{S}$ (with 
$\vec{L}=\vec{x}\times \vec{P}$) do not change their form. Here 
the operators $K^i\equiv X_{(0i)}$ are the analogous of the 
basis-generators of the Lorentz boosts of $SL(2,\Bbb C)$ since in the limit of 
$\omega\to 0$, when $(\ref{mssu})$ equals the Minkowski line element, 
the operators $H,\,P^i,\,J^i$ and $K^i$ become the generators of the spinor 
representation of the group $T(4)\circledS SL(2,\Bbb C)$ (i.e., the universal 
covering group of the Poincar\' e group \cite{WKT,TH}).       

In both these charts the generators (\ref{Gi})-(\ref{Gf}) are self-adjoint 
with respect to the scalar product (\ref{sp}) with the weight function 
(\ref{mu}) if we consider usual boundary conditions on $D\equiv \Bbb R^3$. 
Therefore, for any generator $X$ we have
\begin{equation}\label{adj}
\left<X\psi,\psi'\right>=\left<\psi,X\psi'\right>
\end{equation}
if (and only if) $\psi$ and $\psi'$ are solutions of the Dirac equation which 
behave as  tempered distributions or square integrable spinors with respect to 
(\ref{sp}). Moreover, all these operators commute with $E_D$ as well as any 
other operator from the algebra freely generated by them. We get thus a large 
collection of observables among which we can choose suitable sets of commuting 
operators defining the fermion quantum modes at the level of the relativistic 
quantum mechanics.

\section{Polarized plane waves solutions}

As suggested in \cite{SHI}, the plane wave solutions of the Dirac equation 
with $m\not =0$ must be eigenspinors of the momentum operators $P^i$ 
corresponding to the eigenvalues $p^i$, with the same time modulation as the 
spherical waves. Therefore, we have to look for particular solutions in the 
chart $\{t_c,\vec{x}\}$ involving either  positive frequency plane waves or 
negative frequency ones. Bearing in mind that finally these must be related 
among themselves through the charge conjugation, we assume that, in the 
standard representation of the Dirac matrices (with diagonal $\gamma^0$ 
\cite{TH}), these have the form  
\begin{eqnarray}
\psi^{(+)}_{\vec{p}}&=&  \left(
\begin{array}{c}
f^{+}(t_{c}) \,\alpha(\vec{p})\\
g^{+}(t_{c})\,
\frac{\textstyle \vec{\sigma}\cdot\vec{p}}{\textstyle p}
\,\alpha(\vec{p})
\end{array}\right)
e^{i\vec{p}\cdot\vec{x}} \label{psi+}\\
\psi^{(-)}_{\vec{p}}&=&  \left(
\begin{array}{c}
g^{-}(t_{c})\,
\frac{\textstyle \vec{\sigma}\cdot\vec{p}}{\textstyle p}
\,\beta(\vec{p})\\
f^{-}(t_{c}) \,\beta(\vec{p})
\end{array}\right)
e^{-i\vec{p}\cdot\vec{x}}\label{psi-}
\end{eqnarray}    
where $p=|\vec{p}|$, $\sigma_i$ denotes the Pauli matrices while $\alpha$ and 
$\beta$ are arbitrary Pauli spinors depending on $\vec{p}$. Replacing these 
spinors in the Dirac equation given by (\ref{ED1}) and denoting $k=m/\omega$ 
and $\nu_{\pm}=\frac{1}{2}\pm ik$, we find  equations of the form (\ref{H2}) 
whose solutions can be written in terms of Hankel functions as 
\begin{eqnarray}
&&f^{+}= (-f^{-})^{*}=C{t_{c}}^{2}e^{\pi k/2}H^{(1)}_{\nu_{-}}(-p t_{c}) 
 \label{fg1}\\ 
&&g^{+}= (-g^{-})^{*}=C{t_{c}}^{2} e^{-\pi k/2}H^{(1)}_{\nu_{+}}(-p t_{c})\,. 
\label{fg2} 
\end{eqnarray}  
The integration constant $C$ will be calculated from the ortonormalization 
condition  in the momentum scale. 
 
Hence the plane wave solutions are determined up to the significance of the 
Pauli spinors $\alpha$ and $\beta$. For $\vec{p}\not = 0$ these can be treated 
as in the flat case \cite{SW1,TH} since, in the tetrad gauge (\ref{tt}), the 
spaces of these spinors carry unitary linear representations of the 
$\tilde E(3)$ group. Indeed, a transformation (\ref{es}) produced by 
$(A, \phi_{A,\vec{a}})\in \tilde E(3)\subset S(M)$ with $A\in SU(2)$ 
and $\vec{a}\in {\Bbb R}^3$, involves the linear isometry 
$\vec{x}\to {\vec{x}}^{\,\prime}=\Lambda(A)\vec{x}+\vec{a}$ and the global 
transformation $\psi(t,\vec{x})\to \psi^{\prime}(t,{\vec{x}}^{\,\prime})= 
\rho(A)\psi(t,\vec{x})$. Consequently, the Pauli spinors 
transform according to the  unitary (linear) representation  
\begin{equation}\label{tri}
\alpha(\vec{p})\to e^{-i\vec{a}\cdot\vec{p}}\,A\,\alpha[\Lambda(A)^{-1}
\vec{p}\,]
\end{equation}
(and similarly for $\beta$) that conserves the orthogonality. 
This means that any pair of orthogonal spinors, 
$\tilde\xi_{\sigma}(\vec{p}),\,  \sigma=\pm 1/2$, 
(obeying $\tilde \xi_{\sigma}^{+}\tilde
\xi_{\sigma '}=\delta_{\sigma,\sigma'}$), represents 
an arbitrary basis in the space of Pauli spinors 
\begin{equation}\label{alpha}
\alpha(\vec{p})= \sum_{\sigma} \tilde \xi_{\sigma}(\vec{p}) a(\vec{p},\sigma) 
\end{equation}   
whose components, $a(\vec{p}, \sigma)$, are the wave functions 
in momentum representation of a particle of momentum $\vec{p}$ and  
polarization $\sigma$. 
According to the standard interpretation of the negative frequency terms 
\cite{SW1,TH}, the corresponding basis of the space of $\beta$ spinors is 
defined by
\begin{equation} \label{beta}
\beta(\vec{p})= \sum_{\sigma} \tilde \eta_{\sigma}(\vec{p}) 
[b(\vec{p},\sigma)]^{*}
\,, \quad
\tilde \eta_{\sigma}(\vec{p})=
i\sigma_2 [\tilde \xi_{\sigma}(\vec{p})]^{*} 
\end{equation}   
where $b(\vec{p},\sigma)$ are the antiparticle wave functions. It remains to 
choose the concrete basis we need, using supplementary physical assumptions. 
Since it is not sure that the so called spin basis \cite{SW1} can be 
correctly defined in de Sitter geometry, we prefer the {\em helicity} basis. 
This is formed by the orthogonal Pauli spinors of helicity 
$\lambda=\pm 1/2$ which fulfill 
\begin{equation}\label{heli}
\vec{\sigma}\cdot\vec{p}\,\,\tilde\xi_{\lambda}(\vec{p})=2p\lambda\,
\tilde\xi_{\lambda}(\vec{p})    
\,, \quad 
\vec{\sigma}\cdot\vec{p}\,\,\tilde\eta_{\lambda}(\vec{p})=-2p\lambda\,
\tilde\eta_{\lambda}(\vec{p})\,.    
\end{equation}

The desired particular solutions of the Dirac equation with $m\not=0$ result 
from (\ref{psi+}) and (\ref{psi-}) where we replace (\ref{fg1}) and (\ref{fg2}) 
and the spinors (\ref{alpha}) and (\ref{beta}) written in the 
helicity basis (\ref{heli}). It remains to calculate the normalization 
constant $C$ with respect to the scalar product (\ref{sp}) with the weight 
function (\ref{mu}). After few manipulations in the chart $\{t,\vec{x}\}$, it 
turns out the final form of the fundamental spinor solutions of positive and 
negative frequencies with momentum $\vec{p}$ and helicity $\lambda$,   
\begin{eqnarray}
U_{\vec{p},\lambda}(t,\vec{x})&=& i N\left(
\begin{array}{c}
\frac{1}{2}\,e^{\pi k/2}H^{(1)}_{\nu_{-}}(qe^{-\omega t}) \,
\tilde\xi_{\lambda}(\vec{p})\\
\lambda\, e^{-\pi k/2}H^{(1)}_{\nu_{+}}(qe^{-\omega t})
\,\tilde\xi_{\lambda}(\vec{p})
\end{array}\right)
e^{i\vec{p}\cdot\vec{x}-2\omega t}\label{Ups}\\
V_{\vec{p},\lambda}(t,\vec{x})&=&iN  \left(
\begin{array}{c}
-\lambda\,e^{-\pi k/2}H^{(2)}_{\nu_{-}}(qe^{-\omega t})\,
\tilde\eta_{\lambda}(\vec{p})\\
\frac{1}{2}\,e^{\pi k/2}H^{(2)}_{\nu_{+}}(qe^{-\omega t}) \,\tilde\eta_{\lambda}(\vec{p})
\end{array}\right)
e^{-i\vec{p}\cdot\vec{x}-2\omega t}\,,\label{Vps}
\end{eqnarray}    
which depend on the new parameter $q=p/\omega$ and 
\begin{equation}
N=\frac{1}{(2\pi)^{3/2}}\sqrt{\pi q}\,. 
\end{equation}
Using (\ref{H1}) and (\ref{H3}), it is not hard to verify that 
these spinors are charge-conjugated to each other,   
\begin{equation}\label{conj}
V_{\vec{p},\lambda}=(U_{\vec{p},\lambda})^{c}=C
(\overline{U}_{\vec{p},\lambda})^T \,, \quad C=i\gamma^2\gamma^0\,,   
\end{equation}
satisfy the ortonormalization relations
\begin{eqnarray}
&&\left<U_{\vec{p},\lambda},U_{\vec{p}^{\,\prime},\lambda^{\prime}}\right>=
\left<V_{\vec{p},\lambda},V_{\vec{p}^{\,\prime},\lambda^{\prime}}\right>=
\delta_{\lambda\lambda^{\prime}}\delta^3 (\vec{p}-\vec{p}^{\,\prime})\,, 
\label{orto1}\\
&&\left<U_{\vec{p},\lambda},V_{\vec{p}^{\,\prime},\lambda^{\prime}}\right>=
\left<V_{\vec{p},\lambda},U_{\vec{p}^{\,\prime},\lambda^{\prime}}\right>=
0\,,\label{orto2}
\end{eqnarray}
and represent a {\em complete} set of solutions in the sense that
\begin{equation}\label{compl}
\int d^3 p \sum_{\lambda}\left[
U_{\vec{p},\lambda}(t,\vec{x})U^{+}_{\vec{p},\lambda}(t,\vec{x}^{\,\prime})+
V_{\vec{p},\lambda}(t,\vec{x})V^{+}_{\vec{p},\lambda}(t,\vec{x}^{\,\prime})
\right]=e^{-3\omega t}\delta^3 (\vec{x}-\vec{x}^{\,\prime})\,.
\end{equation}
Let us observe that the factor $e^{-3\omega t}$ does not 
have a special physical meaning since its role is only to compensate the 
weight function (\ref{mu}). Other important properties are   
\begin{eqnarray}
P^i\,U_{\vec{p},\lambda}= p^i\, U_{\vec{p},\lambda}\,, &\quad& 
P^i\,V_{\vec{p},\lambda}=-p^i\, V_{\vec{p},\lambda} \,,\label{PUV}\\ 
W\,U_{\vec{p},\lambda}= p\lambda U_{\vec{p},\lambda}\,, &\quad& 
W\,V_{\vec{p},\lambda}=-p\lambda V_{\vec{p},\lambda}\,.\label{WUV} 
\end{eqnarray} 
where $W=\vec{J}\cdot \vec{P}=\vec{S}\cdot\vec{P}$ is the  helicity operator 
which is analogous to the time-like 
component of the four-component Pauli-Lubanski operator of the Poincar\' e  
algebra \cite{WKT}. Thus, we arrive at the conclusion that the 
fundamental solutions (\ref{Ups}) and (\ref{Vps}) form a complete set of common 
eigenspinors of the operators $P^i$ and $W$.    

In the  case of $m=0$ (when $k=0$) it is convenient to consider the chiral 
representation of the Dirac matrices (with diagonal $\gamma^5$ \cite{SW1}) 
and the chart $\{t_c,\vec{x}\}$. We find that the fundamental solutions 
in helicity basis of the left-handed massless Dirac field,  
\begin{eqnarray}
U^0_{\vec{p},\lambda}(t_c,\vec{x})&=&
\lim_{k\to 0} \frac{1-\gamma^5}{2} U_{\vec{p},\lambda}(t_c,\vec{x})
\nonumber\\
&=&\left(\frac{-\omega t_{c}}{2\pi}\right)^{3/2}
\left(
\begin{array}{c}
(\frac{1}{2}-\lambda)\tilde\xi_{\lambda}(\vec{p})\\
0
\end{array}\right)
\,e^{-ipt_{c}+i\vec{p}\cdot\vec{x}} \label{n1}\\ 
V^0_{\vec{p},\lambda}(t_c,\vec{x})&=&
\lim_{k\to 0} \frac{1-\gamma^5}{2}V_{\vec{p},\lambda}(t_c,\vec{x})
\nonumber\\
&=&\left(\frac{-\omega t_{c}}{2\pi}\right)^{3/2}
\left(
\begin{array}{c}
(\frac{1}{2}+\lambda)\tilde\eta_{\lambda}(\vec{p})\\
0
\end{array}\right)
\,e^{ipt_{c}-i\vec{p}\cdot\vec{x}}\,,  \label{n2}
\end{eqnarray}  
are non-vanishing only for positive frequency and $\lambda=-1/2$ or 
negative frequency and $\lambda=1/2$, as in Minkowski spacetime.
Obviously, these solutions have similar properties as 
(\ref{conj})-(\ref{WUV}).

\section{Quantization}

The quantization can be done simply considering that the wave functions 
in momentum representation, $a(\vec{p},\lambda)$ and $b(\vec{p},\lambda)$, 
become field operators 
(so that $b^{*}\to b^{\dagger}$). Then the quantum  field which satisfies 
the Dirac equation with $m\not=0$ in the chart $\{t,\vec{x}\}$ reads    
\begin{eqnarray}
\psi(t,\vec{x})&=&\psi^{(+)}(t,\vec{x})+\psi^{(-)}(t,\vec{x})\nonumber\\
&=&\int d^3 p \sum_{\lambda}\left[U_{\vec{p},\lambda}(x)
a(\vec{p},\lambda)+V_{\vec{p}, \lambda}(x)b^{\dagger}(\vec{p}, \lambda)
\right]\,.\label{psiab}
\end{eqnarray}
We believe that the particle ($a$, $a^{\dagger}$) and antiparticle 
($b$, $b^{\dagger}$) operators must fulfill the standard anticommutation 
relations in the momentum representation, from which the non-vanishing ones 
are   
\begin{equation}\label{acom}
\{a(\vec{p},\lambda), a^{\dagger}({\vec{p}}^{\,\prime},\lambda^{\prime})\}=
\{b(\vec{p},\lambda), b^{\dagger}({\vec{p}}^{\,\prime},\lambda^{\prime})\}=
\delta_{\lambda,\lambda^{\prime}}\delta^3 (\vec{p}-{\vec{p}}^{\,\prime})\,, 
\end{equation}
since then the equal-time anticommutator takes the {\em canonical} form
\begin{equation}
\{ \psi(t,\vec{x}),\, \overline{\psi}(t, \vec{x}^{\,\prime})\}=
e^{-3\omega t}\gamma^0 \delta^{3}(\vec{x}-\vec{x}^{\,\prime})\,, 
\end{equation}
as it results from (\ref{compl}). In general, the partial anticommutator 
functions,
\begin{equation}
\tilde S^{(\pm)}(t,t',\vec{x}-\vec{x}^{\,\prime})=
i\{ \psi^{(\pm)}(t,\vec{x}),\, 
\overline{\psi}^{(\pm)}(t', \vec{x}^{\,\prime})\}\,,
\end{equation}
and the total one $\tilde S=\tilde S^{(+)}+\tilde S^{(-)}$ are rather 
complicated since for $t\not=t'$ we have no more identities like (\ref{H3}) 
which should simplify their time-dependent parts. In any event, these are 
solutions of the Dirac equation in both their sets of coordinates and help one 
to write the Green functions in usual manner. For example, from the standard 
definition of the Feynman propagator \cite{SW1},  
\begin{eqnarray}
&&\tilde S_F(t,t',\vec{x}-\vec{x}^{\,\prime})=
i\left<0\right|T[\psi(x)\overline{\psi}(x')]\left|0\right>\\
&&=\theta(t-t')\tilde S^{(+)}(t,t',\vec{x}-\vec{x}^{\,\prime})-
\theta(t'-t)\tilde S^{(-)}(t,t',\vec{x}-\vec{x}^{\,\prime})\,, 
\end{eqnarray} 
we find that 
\begin{equation}
E_{D}(x)\tilde S_F(t,t',\vec{x}-\vec{x}^{\,\prime})=
-e^{-3\omega t} \delta^{4}(x-x')\,.
\end{equation}  

Another argument for this quantization procedure is that the one-particle 
operators given by the Noether theorem  have similar structures and 
properties like those of the quantum theory of the free fields in flat 
spacetime. Indeed, according to (\ref{cq}) and (\ref{adj}), the Noether 
theorem guarantees that for any generator $X$ of the spinor representation of 
$S(M)$ there exists a {\em conserved} one-particle operator of the quantum 
field theory which can be calculated simply as 
\begin{equation}\label{opo}
{\bf X}=:\left<\psi, X\psi\right>: 
\end{equation}
respecting the normal ordering of the operator products \cite{SW1}. Hereby
we recover the standard algebraic properties
\begin{equation}
[{\bf X}, \psi(x)]=-X\psi(x)\,, \quad 
[{\bf X}, {\bf X}']=:\left<\psi, [X,X']\psi\right>: 
\end{equation}
due to the canonical quantization adopted here. 

The diagonal one-particle operators result directly from 
(\ref{opo})  and (\ref{orto1})-(\ref{WUV}). In this way 
we obtain the momentum components 
\begin{equation}
{\bf P}^i=:\left<\psi,P^{i}\psi\right>:=
\int d^3 p\, p^i\sum_{\lambda}
\left[a^{\dagger}(\vec{p},\lambda)a(\vec{p},\lambda)
+b^{\dagger}(\vec{p},\lambda)b(\vec{p},\lambda)\right] 
\end{equation}
and the Pauli-Lubanski operator
\begin{equation}
{\bf W}=:\left<\psi,W\psi\right>:=  
\int d^3 p  \sum_{\lambda} p\lambda 
\left[a^{\dagger}(\vec{p},\lambda)a(\vec{p},\lambda)
+b^{\dagger}(\vec{p},\lambda)b(\vec{p},\lambda)\right]\,.
\end{equation}
The definition (\ref{opo}) holds for the generators of internal symmetries too, 
including the particular case of  $X=1$ when the bracket
\begin{equation}
{\bf Q}=\,:\left<\psi,\psi\right>:\,=  
\int d^3 p  \sum_\lambda \left[a^{\dagger}(\vec{p},\lambda)a(\vec{p},\lambda)
-b^{\dagger}(\vec{p},\lambda)b(\vec{p},\lambda)\right]
\end{equation}
gives just the charge operator corresponding to the internal $U(1)$ symmetry 
of (\ref{action}) \cite{TH,BD}. It is obvious that all these operators 
are generators of the external or internal symmetry transformations of the 
quantum fields \cite{SW1}. The conclusion is that the helicity states of the 
Fock space are common eigenstates of the set $\{{\bf Q}, {\bf P}^i, {\bf W}\}$. 

The Hamiltonian  operator ${\bf H}=:\left<\psi,H\psi\right>:$ is conserved but 
is not diagonal in this basis since it does not commute  with ${\bf P}^i$ and 
${\bf W}$. Its form in momentum representation can be calculated using the 
identity
\begin{equation}
H\,U_{\vec{p},\lambda}(t,\vec{x})=-i\omega
\left(p^i\partial_{p^{i}}+\frac{3}{2}\right)
U_{\vec{p},\lambda}(t,\vec{x}) 
\end{equation}
and the similar one for $V_{\vec{p},\lambda}$, leading to  
\begin{equation}
{\bf H}=\frac{i\omega}{2}\int d^3 p\,p^i\sum_{\lambda}\left[
a^{\dagger}(\vec{p},\lambda)\stackrel{\leftrightarrow}{\partial}_{p^{i}}
a(\vec{p},\lambda) +
b^{\dagger}(\vec{p},\lambda)\stackrel{\leftrightarrow}{\partial}_{p^{i}}
b(\vec{p},\lambda) \right]
\end{equation}
where the derivatives act as $f\stackrel{\leftrightarrow}{\partial}h=f\partial 
h -(\partial f) h$. Hereby it results the expected  behavior of ${\bf H}$ 
under the  space translations of $\tilde E(3)$ which 
transform the operators $a$ and $b$ according to (\ref{tri}). Moreover, it is 
worth pointing out that, there exist momentum dependent phase transformations  
\begin{eqnarray}
U_{\vec{p},\lambda}&\to& e^{i\chi(\vec{p})} U_{\vec{p},\lambda}\label{gaugep} \\ 
V_{\vec{p},\lambda}&\to& e^{-i\chi(\vec{p})} V_{\vec{p},\lambda} \\ 
a(\vec{p},\lambda)&\to&e^{-i\chi(\vec{p})}a(\vec{p},\lambda) \\
b(\vec{p},\lambda)&\to&e^{-i\chi(\vec{p})}b(\vec{p},\lambda) 
\end{eqnarray}
(with $\chi \in \Bbb R$) leaving invariant the operators 
$\psi,\, {\bf Q},\,{\bf P}^{i}$ and ${\bf W}$ as well as the equations 
(\ref{acom}), while  the Hamiltonian operator transforms as
\begin{equation}\label{trah} 
{\bf H}\to {\bf H} +
\omega\int d^3 p\, [p^i \partial_{p^i}\chi(\vec{p})] 
\sum_\lambda \left[a^{\dagger}(\vec{p},\lambda)a(\vec{p},\lambda)
+b^{\dagger}(\vec{p},\lambda)b(\vec{p},\lambda)\right]\,.
\end{equation}
This interesting property may be interpreted as a new type of gauge 
transformations depending on  momentum instead of coordinates. Our
preliminary calculations indicate that this gauge may be helpful for 
analyzing the behavior of our theory near $\omega\sim 0$.  

In the simpler case of the left-handed massless field which has 
the fundamental spinor solutions (\ref{n1}) and (\ref{n2}) we obtain similar 
results among them we recover the standard rule of neutrino polarization.

\section{Concluding remarks}

We have derived here a complete set of normalized plane wave solutions of the 
Dirac equation in the chart with the line element (\ref{mssu}) of the de Sitter 
spacetime. These determine the quantum modes of polarized free fermions  
characterized by  momentum and helicity. Thus we obtain a similar approach 
as in special relativity allowing us to perform the second quantization of the 
free Dirac field according to the canonical method. 

It remains to solve many associated problems from the pure mathematical ones up 
to those regarding to the physical interpretation of the specific observables 
of the de Sitter background. First of all, one has to look for an orbital 
analysis analogous to the Wigner theory of the induced representations of the 
Poincar\' e group \cite{WKT,TH}. This could help one to understand the meaning 
of the rest frames (with $\vec{p}=0$) in the de Sitter geometry indicating 
which are the ``booster" mechanisms giving rise to waves of arbitrary momentum 
from those with $\vec{p}=0$. In the same time it is important to investigate 
the physical consequences of the relations among the Hamiltonian operator and 
the other generators of the spinor representation of the $S(M)$ group and the 
role and significance of the gauge transformations (\ref{gaugep})-(\ref{trah}). 
Moreover, in further investigations of the quantum free Dirac field, some 
specific problems could appear but these seem to be rather technical, e.g. the 
properties of commutator and Green functions, calculation of the action of 
more complicated conserved operators, evaluation of the inertial effects etc. 
However, in our opinion, the next important step would be to construct a 
similar theory for the free electromagnetic field which should complete the 
basic ingredients one needs for developing the perturbative QED in de Sitter 
spacetime.

The results  obtained here show that, even though many particular features of 
the quantum theory in curved spacetimes depend on the choice of the local chart 
and tetrad gauge, there are covariance properties providing us with operators 
with invariant commutation relations. For this reason we hope that our approach  
based on external symmetries could be an argument for a general tetrad gauge 
covariant theory of quantum fields with spin in which the second quantization 
should be independent on the frames where one works.

\appendix

\section{SO(4,1) transformations and isometries}

The spacetime $M^5$ is pseudo-Eucidean with the metric $\eta^5$ that is 
invariant under the coordinate transformations $Z^A\to {^5\!\Lambda}^{A\,\cdot}
_{\cdot\,B}Z^B$ where $^5\!\Lambda\in SO(4,1)$. Each coordinate transformation 
give rise to an isometry of $M$ which can be calculated in the local chart 
$\{t_{c},\vec{x}\}$ using the equations (\ref{Zx}). We remind the reader that 
the basis generators $^5\! X_{AB}$ of the fundamental (linear) representation of 
$SO(4,1)$, carried by $M^5$, have the matrix elements
\begin{equation}
(^5\! X_{AB})^{C\,\cdot}_{\cdot\,D}=i\left(\delta^C_A\, \eta_{BD} 
-\delta^C_B\, \eta_{AD}\right)\,.
\end{equation} 

The transformations of $SO(3)\subset SO(4,1)$ are simple rotations of $Z^i$ 
and $x^i$ which transform alike since this symmetry is global. For the other 
transformations generated by $H,\, P^i$ and $N^i$ the linear transformations 
in $M^5$ and the isometries are different. Those due to $H$, 
\begin{equation}
e^{-i\xi_{H}{^5\!H}}\,:\quad
\begin{array}{lcl}
Z^0&\to&Z^0 \cosh\alpha-Z^5 \sinh\alpha \\   
Z^5&\to&-Z^5 \sinh\alpha+Z^0 \cosh\alpha \\   
Z^i&\to&Z^i     
\end{array} 
\end{equation}
where $\alpha=\omega\xi_{H}$, produce the dilatations
$t_{c}\to t_{c}e^{\alpha}$ and $x^i\to x^i e^{\alpha}$,
while the transformations
\begin{equation}
e^{-i\vec{\xi}_{P}\cdot {^5\!\vec{P}}}\,:\quad
\begin{array}{lcl}
Z^0&\to&Z^0 +\omega\,\vec{\xi}_{P}\cdot\vec{Z}+\frac{1}{2}\,\omega^2 
{{\xi}_P}^2\,(Z^0+Z^5) \\   
Z^5&\to&Z^5 -\omega\,\vec{\xi}_{P}\cdot\vec{Z}-\frac{1}{2}\,\omega^2 
{{\xi}_P}^2\,(Z^0+Z^5) \\   
Z^i&\to&Z^i+\omega\,\xi_P^i\,(Z^0+Z^5)    
\end{array}
\end{equation}
give the space translations $ x^i\to x^i +\xi_P^i$ at fixed $t_c$. 
More interesting are the transformations
\begin{equation}
e^{-i\vec{\xi}_{N}\cdot {^5\!\vec{N}}}\,:\quad
\begin{array}{lcl}
Z^0&\to&Z^0 -\vec{\xi}_{N}\cdot\vec{Z}+\frac{1}{2}\, 
{{\xi}_N}^2\,(Z^0-Z^5) \\   
Z^5&\to&Z^5 -\vec{\xi}_{N}\cdot\vec{Z}+\frac{1}{2}\, 
{{\xi}_N}^2\,(Z^0-Z^5) \\   
Z^i&\to&Z^i-\xi_N^i\,(Z^0-Z^5)    
\end{array}
\end{equation}
which lead to the isometries
\begin{eqnarray}
t_c&\to&\frac{t_c}{1-2\omega\, \vec{\xi}_{N}\cdot\vec{x} 
-\omega^2{{\xi}_N}^2\,({t_c}^2-r^2)} \\
x^i&\to&\frac{x^i+\omega\xi_N^i\, ({t_c}^2-r^2)}
{1-2\omega\, \vec{\xi}_{N}\cdot\vec{x} 
-\omega^2{{\xi}_{N}}^2\,({t_c}^2-r^2)}\,.
\end{eqnarray}
We denoted here ${\xi_P}^2=(\vec{\xi}_P)^2$ and ${\xi_N}^2=(\vec{\xi}_N)^2$.  

\section{Some properties of Hankel functions}

According to the general properties of the Hankel functions \cite{AS}, we 
deduce that those used here, $H^{(1,2)}_{\nu_{\pm}}(z)$, with 
$\nu_{\pm}=\frac{1}{2}\pm i k$ and $z\in \Bbb R$, are related among 
themselves through 
\begin{equation}\label{H1}
[H^{(1,2)}_{\nu_{\pm}}(z)]^{*} 
=H^{(2,1)}_{\nu_{\mp}}(z)\,,  
\end{equation}
satisfy the equations
\begin{equation}\label{H2} 
\left(\frac{d}{dz}+\frac{\nu_{\pm}}{z}\right)H^{(1,2)}_{\nu_{\pm}}(z) 
=  i e^{\pm \pi k} H^{(1,2)}_{\nu_{\mp}}(z)
\end{equation}
and the identities
\begin{equation}\label{H3}
e^{\pm \pi k}
H^{(1)}_{\nu_{\mp}}(z)H^{(2)}_{\nu_{\pm}}(z)
+ e^{\mp \pi k} H^{(1)}_{\nu_{\pm}}(z)H^{(2)}_{\nu_{\mp}}(z)=\frac{4}{\pi z}\,.
\end{equation}

%\end{document}

\end{document}